\documentclass[aps,prapplied,preprint,longbibliography,superscriptaddress]{revtex4-2}
\usepackage{graphicx} 
\usepackage{amsmath,amssymb}
\usepackage{booktabs}
\usepackage{array}
\usepackage{xcolor}
\usepackage{wrapfig}
\usepackage{enumitem}
\usepackage{natbib}
\usepackage{tabularx}
\usepackage[colorlinks=true,
            linkcolor=blue,
            citecolor=blue,
            urlcolor=blue]{hyperref}

\begin{document}

\title{Fabrication of nano-diamonds with a single {NV} center: Towards matter-wave interferometry with massive objects}

\author{Menachem Givon}
\affiliation{Department of Physics, Ben-Gurion University of the Negev, Beer-Sheva, Israel}

\author{Yaniv Bar-Haim}
\affiliation{Department of Physics, Ben-Gurion University of the Negev, Beer-Sheva, Israel}

\author{David Groswasser}
\affiliation{Department of Physics, Ben-Gurion University of the Negev, Beer-Sheva, Israel}

\author{Asi Solodar}
\affiliation{Nano-Fabrication center, Ben-Gurion University of the Negev, Beer-Sheva, Israel}

\author{Nadav Aharon}
\affiliation{Nano-Fabrication center, Ben-Gurion University of the Negev, Beer-Sheva, Israel}

\author{Michael Belman}
\affiliation{Nano-Fabrication center, Ben-Gurion University of the Negev, Beer-Sheva, Israel}

\author{Amit Yosefi}
\affiliation{Nano-Fabrication center, Ben-Gurion University of the Negev, Beer-Sheva, Israel}

\author{Erez Golan}
\affiliation{Nano-Fabrication center, Ben-Gurion University of the Negev, Beer-Sheva, Israel}

\author{J\"{u}rgen Jopp}
\affiliation{Ilse Katz Institute for Nanoscale Science \& Technology, Ben-Gurion University of the Negev, Beer-Sheva, Israel}
\author{Ron Folman}
\affiliation{Department of Physics, Ben-Gurion University of the Negev, Beer-Sheva, Israel}
\affiliation{Ilse Katz Institute for Nanoscale Science \& Technology, Ben-Gurion University of the Negev, Beer-Sheva, Israel}
\date{18 August 2025}

\begin{abstract}

{Q}uantum mechanics (QM) and {G}eneral relativity (GR), also known as the theory of gravity, are the two pillars of modern physics. A matter-wave interferometer with a massive particle can test numerous fundamental ideas, including the spatial superposition principle---a foundational concept in QM---in previously unexplored regimes. It also opens the possibility of probing the interface between QM and GR, such as testing the quantization of gravity. Consequently, there exists an intensive effort to realize such an interferometer. While several approaches are being explored, we focus on utilizing nanodiamonds with embedded spins as test particles which, in combination with Stern–Gerlach forces, enable the realization of a closed-loop matter-wave interferometer in space-time. There is a growing community of groups pursuing this path \cite{White_paper_Morley2025}. We are posting this technical note (as part of a series of seven such notes), to highlight our plans and solutions concerning various challenges in this ambitious endeavor, hoping this will support this growing community. Here we discuss the design considerations for a high-precision enhanced-coherence nanodiamond source, review the fabrication processes used to produce nanodiamond pillars measuring $40 \times 65 \times 80\,$nm, summarize the characterization work completed to date, and conclude with an outlook on the remaining steps needed to finalize the source fabrication. We would be happy to make available more details upon request.

\end{abstract}

\maketitle

\section{Introduction}
\label{intro}

Quantum mechanics (QM) is a pillar of modern physics. It is thus imperative to test it in ever-growing regions of the relevant parameter space. A second pillar is General relativity (GR), and as a unification of the two seems to be eluding continuous theoretical efforts, it is just as imperative to experimentally test the interface of these two pillars by conducting experiments in which foundational concepts of the two theories must work in concert.

The most advanced demonstrations of massive spatial superpositions have been achieved by Markus Arndt's group, reaching systems composed of approximately 2,000 atoms\,\cite{Fein2019}. This will surely grow by one or two orders of magnitude in the near future. An important question is whether one can find a new technique that would push the state of the art much further in mass and spatial extent of the superposition. Several paths are being pursued \cite{Romero-Isart_2017,Pino_2018,Weiss2021,Neumeier2024,Kialka2022} and we choose to utilize Stern-Gerlach forces.

The Stern-Gerlach interferometer (SGI) has, in the last decade, proven to be an agile tool for atom interferometry \cite{Amit_2019,dobkowski2025observationquantumequivalenceprinciple,Keil2021}. Consequently, we, as well as others, aim to utilize it for interferometry with massive particles, specifically, nanodiamonds (NDs) with a single spin embedded in their center \cite{Scala_2013,Wan_2016,margalit2021_OUR_intro}.
Levitating, trapping and cooling of massive particles, most probably a prerequisite for interferometry with such particles, has been making significant progress in recent years. Specifically, the field of levitodynamics is a fast growing field \cite{gonzalez-ballestero2021_4inOpr}. Commonly used particles are silica spheres. As the state of the art spans a wide spectrum of techniques, achievements and applications, instead of referencing numerous works, we take, for the benefit of the reader, the unconventional step of simply mapping some of the principal investigators; these include Markus Aspelmeyer, Lukas Novotny, Peter Barker, Kiyotaka Aikawa, Romain Quidant, Francesco Marin, Hendrik Ulbricht and David Moore.. Relevant to this work, a rather new sub-field which is now being developed deals with ND particles, where the significant difference is that a spin with long coherence times may be embedded in the ND. Such a spin, originating from a nitrogen-vacancy (NV) center, could enable the coherent splitting and recombination of the ND by utilizing Stern-Gerlach forces \cite{Wan_2016,Scala_2013,margalit2021_OUR_intro}. This endeavor includes principal investigators such as Tongcang Li, Gavin Morley, Gabriel Hetet, Tracy Northup, Brian D’Urso, Andrew Geraci, Jason Twamley and Gurudev Dutt.

We aim to start with a ND of $10^7$ atoms and extremely short interferometer durations. Closing a loop in space-time in a very short time is enabled by the strong magnetic gradients induced by the current-carrying wires of the atom chip \cite{Keil2016}. Such an interferometer will already enable to test the existing understanding concerning environmental decoherence (e.g., from blackbody radiation), and internal decoherence \cite{Henkel2024}, never tested on such a large object in a spatial superposition. As we slowly evolve to higher masses and longer durations (larger splitting), the ND SGI will enable the community to probe not only the superposition principle in completely new regimes, but in addition, it will enable to test specific aspects of exotic ideas such as the Continuous spontaneous localization hypothesis \cite{Adler_2021,Gasbarri2021}. As the masses are increased, the ND SGI will be able to test hypotheses related to gravity, such as modifications to gravity at short ranges (also known as the fifth force), as one of the SGI paths may be brought in a controlled manner extremely close to a massive object \cite{Geraci2010,Geraci2015,Bobowski_2024,Panda2024}. Once SGI technology allows for even larger masses ($10^{11}$ atoms), we could test the Diósi–Penrose collapse hypothesis \cite{Penrose2014,Penrose_2018,Howl_2019,Tomaz_2024,Bassi_2013} and gravity self-interaction \cite{hatifi2023revealingselfgravitysterngerlachhumptydumpty,Grossardt_2021,Aguiar_2024} (e.g., the Schrödinger-Newton equation). Here starts the regime of active masses, whereby not only the gravitation of Earth needs to be taken into account. Furthermore, it is claimed that placing two such SGIs in parallel will allow probing the quantum nature of gravity \cite{Bose2017_quantum_gravity_witness,Marletto_2017}. This will be enabled by ND SGI, as with $10^{11}$ atoms the gravitational interaction could be the strongest \cite{van_de_Kamp_2020,Schut_2023,Schut_2024}.

Let us emphasize that, although high accelerations may be obtained with multiple spins, we consider only an ND with a single spin as numerous spins will result in multiple trajectories and will smear the interferometer signal. We also note that working with a ND with less than $10^7$ atoms is probably not feasible because of two reasons. The first is that NVs that are closer to the surface than 20\,nm lose coherence, and the second is that at sizes smaller than 50\,nm, the relative fabrication errors become large, and a high-precision ND source becomes beyond reach.

In this note, we present the technical details of a key building block of the ND-based SGI. We first discuss the design considerations for the nanodiamond, followed by an outline of the fabrication methods required. We then present our initial fabrication results and conclude with a roadmap for future developments. This technical note
is part of a series of seven technical notes put on the archive towards the end of August 2025,
including a wide range of required building blocks for a ND SGI.

\section{Motivation}
\label{Sec:Motivation}

When aiming for a spatial ND SGI, a few challenges need to be considered. This work considers the ND itself, its size and shape and the placing of an NV spin at its center. The size is crucial for getting a closed loop in space-time as while the SG force is fixed, the mass of the object determines the accelerations. The shape of the ND is also of paramount importance as it affects the angular moment of inertia as well as dipole forces, and these affect torques, rotational alignment and rotational cooling. The need for rotational cooling has been emphasized in our recent work\,\cite{JaphaFolman2023_Uncertainty_Limit_for_SG}. Unfortunately, commercially available NDs are highly unprecise having a large variance in both size and shape. There is consequently an acute need for a method that would provide high-precision NDs at will. Such a method, requiring nano-meter accuracy, obviously must depend on state-of-the-art fabrication techniques.

Let us briefly note that ND SGI can be performed before a high-recision source for NDs is made. When utilizing commercially available NDs, one will need to load NDs into the levitating trap, one after the other, and characterize them until an adequate ND is found. Once such an ND is found, the same ND will have to be reused in all experimental cycles in order to collect the statistics required for obtaining the SGI signal. Such an approach is clearly feasible, but not very efficient. We are currently starting our ND SGI program with commercially available NDs \cite{archive_of_Peter, archive_of_Omer} while in parallel, as presented in this work, we are making significant progress towards a high-precision source.

NDs etched on a chip will not only provide precision. They will provide a dry UHV-compatible source. As collisions with background gas are another source of decoherence, vacuum levels will have to be very good, and depending on the duration of the interferometry, will likely range from $10^{-8}$ to $10^{-14}$ mbar. Existing sources, typically involving liquids, are incompatible with such UHV needs. The launching of the NDs from the surface of the chip towards the trap where they are levitated and in which the experimental sequence may begin, will be done by a combination of vibrations and electric forces (see our work in\,\cite{Rafael_paper}). We do not use laser-based desorption techniques as we would like to keep the internal ND temperatures low for extended NV spin coherence times, and NDs are known to absorb some of the laser radiation. For the same reason, our initial trap is a Paul trap which is dark (before the SGI we neutralize the NDs \cite{archive_of_Sela}).

Crucially, we have shown in \cite{Maria_and_Yoni_archive_paper} that the exact shape of the ND is important in order to achieve rotational control and cooling. To lift the rotational degeneracy, one may very well need to have different object lengths in all three primary axes. Methods of undercut or brittle layer will be used to reduce the strength of the bond between the ND and the substrate on which it was fabricated. 

As described in the following, the combination of state-of-the-art etching with advanced diamond growth techniques, and in addition the use of high-precision localizers such as TEM and STM, can provide the high accuracy positioning of a single NV at the very center of the ND. This is important for two main reasons: first, a maximal distance to the nearest surface increases the coherence time of the spin. Second, applying a SG force on an NV spin which is not centered will induce unwanted torques. 

Furthermore, as noted in the introduction, we aim to start with ND having about $10^7$ atoms and eventually reach NDs having about $10^{11}$ atoms. The former is the minimal size which ensures a distance of at least $20$\,nm from the surface, a distance sufficient for prolonged spin coherence times, and the latter serves as the required milestone for experiments testing the quantization of gravity. While typical fabrication techniques reach high precision (percentage wise) for large objects of a few hundred nm, achieving high precision for $10^7$ atoms (a diameter of tens of nm) is hard, and in this work we show significant progress in this direction. 

Finally, a crucial element is the cleansing of the surface of the ND. A dirty surface giving rise to patch potentials, not only reduces the spin coherence time as well as increasing spatial decoherence mechanisms, but it also introduces forces that may become dominant relative to gravity when searching for spin-spin entanglement due to gravitational interaction in experiments testing the quantization of gravity \cite{Bose2017_quantum_gravity_witness,Marletto_2017}. Here gravity should be the dominant force interacting between two NDs. The fabrication techniques, which we are developing, will also enable highly efficient surface cleaning. 

A schematic of the eventual ND source is shown in Fig.\,1. Visible are the different object lengths in all three primary axes and the undercut, as well as the delta-doped layer in which the single NV is positioned.

\begin{figure}[htbp]
  \centering
  \includegraphics[trim={50 650 20 60}, clip, width=0.95\linewidth]{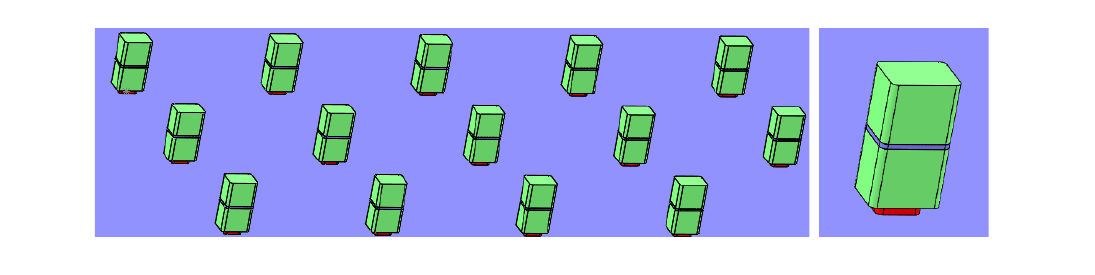} 
  \caption {(Color online) The schematic of an ultra-precision ND source: Left: Part of an array of ND pillars on a diamond substrate (purple). Right: Enlarged single ND pillar. The ND pillar is composed of 3 layers: the two light green layers represent ultra-pure diamond and the middle (blue) layer (1-2 nm thick) represents a delta-doped diamond layer with a small amount of nitrogen. The bottom red layer represents either a brittle layer or an undercut in the ND for easy detachment. Detachment will be achieved by a combination of vibration (with a transducer placed on the chip) and electrical forces \cite{Rafael_paper}. The most general shape of the ND pillar is cuboidal where the height, width and thickness dimensions are all in principle different. Concerning the NV, we have also done considerable work as described in \cite{archive_of_Naor} and references therein.}
  \label{Fig:ND_source_final}
\end{figure}

\section{Design consideration}
\label{Sec:Desing_consideration}
We can broadly divide the design considerations of the ND pillars into two main categories: external and internal. The external aspect includes geometrical parameters such as size and shape, while the internal considerations (which will be discussed in Section \ref{Sec:Outlook}) concern factors related to the NV center. The primary geometrical requirements pertain to the dimensions and volume of the ND pillars. Four key considerations guide their specification:
(i) The distance between the NV center (located at the center of the pillar) and all ND surfaces must exceed $20-30$\,nm to avoid coherence loss due to rapidly fluctuating surface spins \cite{Wang2016Coherence_NV_V_depth,Zhang2017DepthDependent_coherence}. 
(ii) The ND pillar should have a cuboidal shape with distinct length, width, and height to enable independent control over all three rotational degrees of freedom. (This could later be relaxed to an ellipsoid shape if one axis is rotationally accelerated for gyroscopic stabilization \cite{Zhou2025Gyroscopic,Rizaldy2024Rotational,Wachter2025Gyroscopically}.
(iii) The height of the ND pillar should be the largest dimension to facilitate reliable detachment of the pillar from the carrier substrate.
(iv)To test the quantization of graviyt in experiments involving two NDs, the mass of each of them must be at least $10^{-15}\,$kg, corresponding to about $10^{11}$ carbon atoms \cite{Bose2017_quantum_gravity_witness}. Based on these criteria, the ND pillars dimensions are summarized in Table \ref{Table:Dimentions}: $\text{ND}_1$ is the smallest possible ND pillar, $\text{ND}_3$ in the minimal size needed for quantization-of-gravity experiments and $\text{ND}_2$ is an indeterminate size.  

\begin{table}[h!]
\centering
\caption{Properties of different ND pillars}
\label{Table:Dimentions}
\renewcommand{\arraystretch}{1.2}
\begin{tabular}{|l|c|c|c|c|}
\hline
\textbf{} & \textbf{ND}$_1$ & \textbf{ND}$_2$ & \textbf{ND}$_3$ & \textbf{Unit} \\
\hline
Length & 65 & 200 & 900 & nm \\
\hline
Width & 45 & 100 & 400 & nm \\
\hline
Height & 80 & 300 & 1600 & nm \\
\hline
Weight & $7.8\cdot 10^{-19}$ & $2.1\cdot 10^{-17}$ & $2.0\cdot 10^{-15}$ & kg \\
\hline
No. of atoms & $3.9\cdot 10^7$ & $1.1\cdot 10^9$ & $1.0\cdot 10^{11}$ & \\
\hline
\end{tabular}
\end{table}
Finally, the mass uniformity of the fabricated ND pillars must be defined. Many parameters in the interferometry cycle depend on the pillar's mass, which is measured prior to the experiment (see \cite{Ricci2019_Mass_Measurement}). When an ND pillar is replaced, its mass must be remeasured (e.g., to make sure what was trapped is not a cluster). However, as noted, high mass uniformity across the pillars allows for a much more efficient experimental cycle. Our goal is to achieve a standard deviation in the pillars' mass of less than 2\% of the average mass of all the pillars on a particular substrate. To the best of our knowledge, this is beyond the state of the art for our smallest pillars (ND$_1$).

\vspace{1cm}

\section{Fabrication}
\label{Sec:fabrication}
The fabrication of the ND pillar source combines two complementary technologies. First, a bottom-up growth process produces an ultra-pure diamond plate with single NV centers positioned at predetermined locations (see Section \ref{Sec:Outlook}). This is followed by a top-down etching process that defines ND pillars, each centered on an individual NV center. This section, together with Section \ref{Sec:Methods}, describes the key aspects of the top-down fabrication process for the pillars listed in Table \ref{Table:Dimentions}.

\begin{figure}[htbp!]
  \centering
  \includegraphics[trim={0 0 0 0}, clip,width=0.6\linewidth]{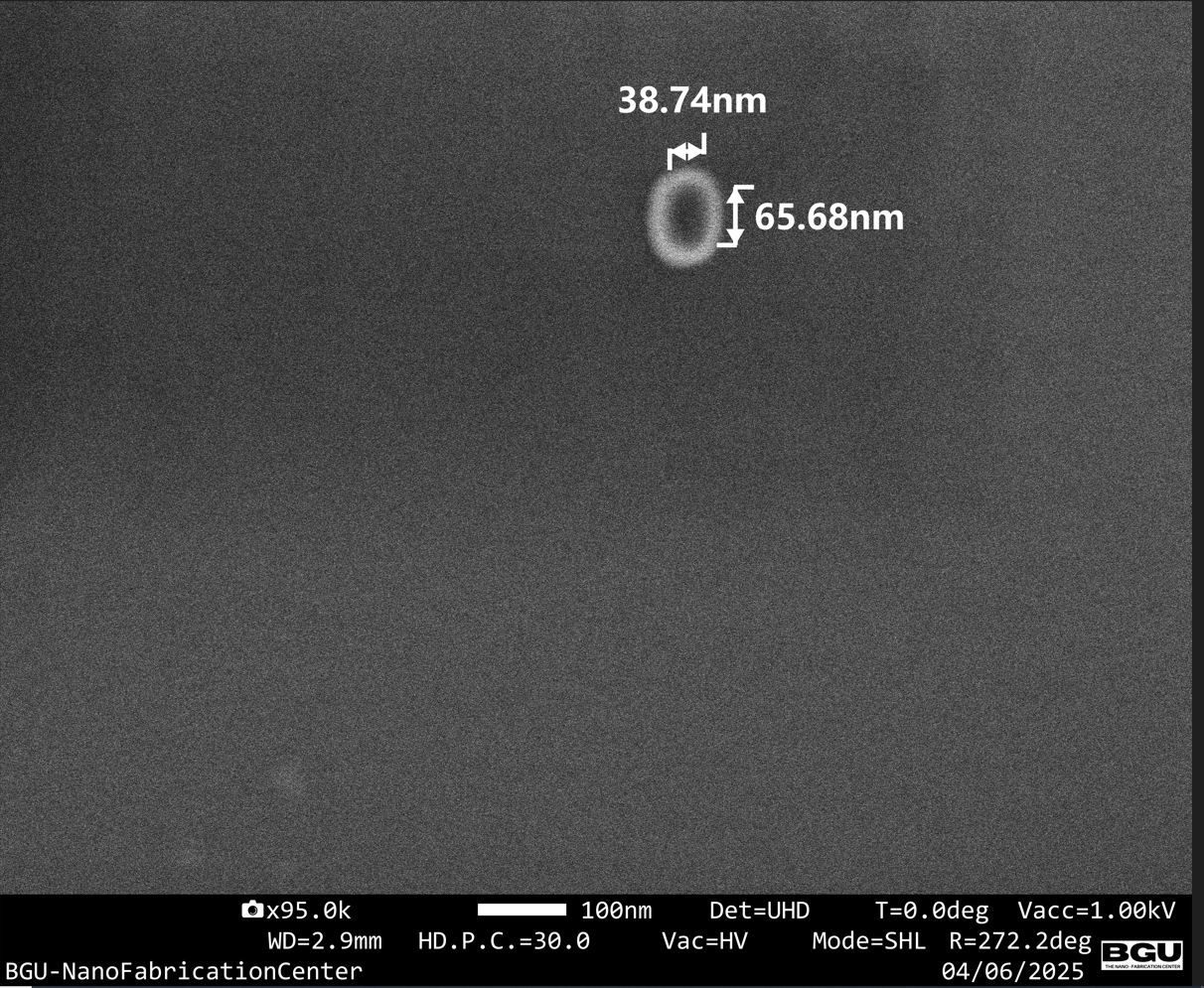} 
  \caption {(Color online) Top-view Scanning Electron Microscope  (SEM) image showing a fabricated single ND pillar (nominal dimensions $40 \times 65\,$nm).}
  \label{Fig:40x55_single}
\end{figure}

Several fabrication methods for diamond nanostructures with dimensions comparable to those of our ND$_1$ pillars have been reported. Top-down approaches based on self-guided reactive ion etching \cite{Trusheim2014_selfguided} or the use of self-assembled masks (see \cite{self_assembly_top_down} and references therein) can produce ND pillars of similar size. However, self-assembly techniques are not compatible with our requirements because the etched ND positions must be aligned with predetermined NV center locations. A deterministic top-down approach \cite{Mobile_probe_nanodiamond_Awschalom} yields much larger NDs, which could be suitable for fabricating ND$_2$ and ND$_3$ structures (Table \ref{Table:Dimentions}). Other works \cite{3D_hollow_nanostructures,Jonker2021_3D+_hollow} demonstrate the fabrication of nanostructures with dimensions similar to ND$_1$, but the dimensional uncertainty in these methods typically ranges from 10\% to 20\%, which is significantly higher than the precision required for our application.

We reviewed several additional fabrication approaches \cite{Paras_Low-Dim_nanomaterials}, including the use of a novel electron beam resist (Medusa 84 SiH by Allresist \cite{Medusa}). Following a series of trials performed on $2.6\times2.6\times0.25\,$mm diamond substrates, we developed the fabrication process described in Subsection \ref{Sec:Fab_Methods}.

\begin{figure}[htbp!]
  \centering
   \includegraphics[trim={70 250 60 50}, clip,width=0.99\linewidth]{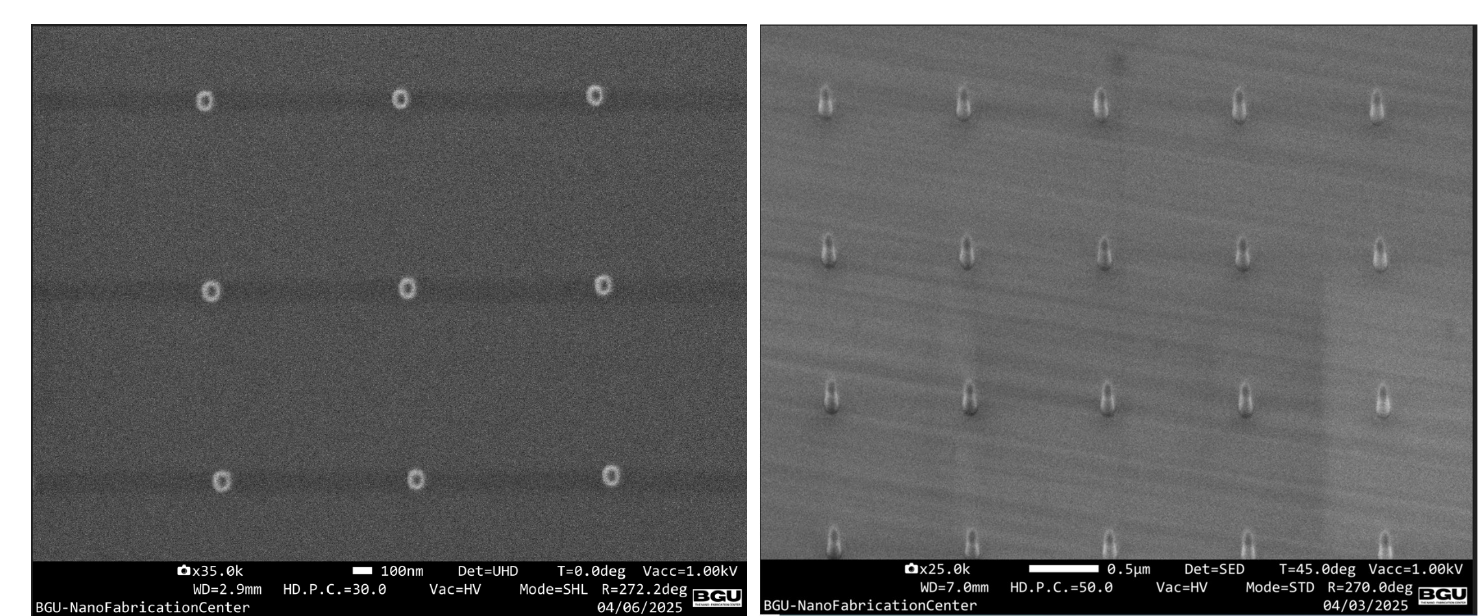} 
   \caption {(Color online) Partial SEM images of an array of ND pillars with lateral dimensions of $40 \times 65\,$nm. Left: Top view. Right: Tilted (45°) view.}
   \label{Fig:40x65_several}
\end{figure}

Figure \ref{Fig:40x55_single} shows a single ND pillar fabricated with nominal dimensions $40 \times 65\,$nm.  Arrays of such pillars are partially shown in Fig.\,\ref{Fig:40x65_several}. Additional pillars with varying dimensions were also fabricated on the same diamond substrate. For instance, a section of an array consisting of $250 \times 400\,$nm pillars is shown in the Subsection \ref{Sec:validation}.     

Characterization of the fabricated ND pillars includes height measurements and area estimation. For a batch of 25 pillars, the average height, as measured with an AFM, is 68.2\thinspace nm with a standard deviation of 0.5\thinspace nm (see Subsection \ref{Sec:Height_Methods}). This corresponds to a relative standard deviation (RSD)---defined as the standard deviation divided by the average and expressed as a percentage---of 0.7\%. This level of variation is considered well within acceptable limits. 

The homogeneity of the areas of the ND pillars depends strongly on their size. Using the normalized area estimation method (see Subsection \ref{Sec:Measure_Methods}) we find an RSD of 1\% for a batch of 60 pillars with dimensions  $250 \times 400\,$nm pillars. In contrast, for a batch of 440 smaller pillars ($40 \times 65\,$nm), the RSD increases to 5\% which may be too high.

This observed variance may stem from two sources: (i) actual fluctuations in the fabricated pillar areas, and (ii) variation introduced by the measurement method. To estimate the latter, we prepared a ``master target" (see Subsection \ref{Sec:Master_target}) and evaluated the measurement-induced variance (see Subsection \ref{Sec:validation}). The results suggest that the measurement process itself can introduce an upper limit of 3–5\% variance, which is too large to reliably isolate the true fabrication-induced variation.

The main fabrication procedures and corresponding results are outlined below.
\begin{itemize}
\item Fabrication of ND$_1$ pillars (Table \ref{Table:Dimentions}) , see Subsection \ref{Sec:Fab_Methods}. We achieved a height uniformity of (RSD) 0.7\%,  cross-section uniformity (RSD) $<5\%$.
\item Fabrication of larger pillars, see  Subsection \ref{Sec:Fab_Methods} and Fig. \ref{Fig:large_structures}. We achieved a height uniformity of (RSD) 0.7\%,  cross-section uniformity (RSD) $<1\%$.
\item Fabrication method for pillars of type ND$_2$ and ND$_3$ (Table \ref{Table:Dimentions}), see Subsection \ref{Sec:ND2_ND3}.   
\item Estimation method for pillars height, see Subsection \ref{Sec:Height_Methods}. We achieved an RSD below 0.7\%. 
 \item Estimation method for pillars' cross-sectional area, see Subsections \ref{Sec:Measure_Methods}.
 \item Fabrication of a master target (Subsection \ref{Sec:Master_target}) for the validation of the cross-sectional estimation method (Subsection \ref{Sec:validation}). We achieved an RSD of 3-5\% for structures similar to ND$_1$ pillars.  More work is needed to reduce it to an RSD below 1\% . 
\item Undercut fabrication method is outlined in Subsection \ref{Sec:Undercut}.  
\end{itemize}

\section{Discussion}

As noted, the ND SGI experiment can be performed with currently available commercial NDs. However, realizing an ultra-precision source of NDs will be highly advantageous. We have therefore invested considerable resources in this goal, and have made significant progress. In our most challenging task of fabricating very small NDs, we have a achieved an accuracy of a few percent. Just as important is our ability to design different shapes which allow us to address theoretically-based requirements regarding rotation alignment and rotation cooling, as well as needs originating in ideas such as gyroscopic stabilization. Taking into account that all required technologies have been realized and what is now required is simply to integrate them into one process, we estimate the source to be available in the near future.

\section{The NV center}
\label{Sec:Outlook}

We have already done considerable work regarding the manipulation of the NV spin, as described in \cite{archive_of_Naor}. Here we focus on fabrication aspects concerning the NV center.

In the opening paragraph of Section\,\ref{Sec:fabrication}, we noted that fabrication of the ND pillar source (schematically shown in Fig.\ref{Fig:ND_source_final}) relies on two complementary approaches: a bottom-up process and a top-down process. The bottom-up process produces two ultra-pure diamond plates with a thin delta-doped layer between them, in which a single NV center is embedded. The top-down process defines the ND pillars by etching, as described in Section\,\ref{Sec:fabrication}. In what follows, we focus on the key aspects of the bottom-up fabrication.

The bottom-up part of the ND pillar source starts with growing an ultra-pure single crystal diamond layer on a high-quality diamond substrate. The thickness of this layer is half the height of the type of pillar we intend to fabricate (see Table \ref{Table:Dimentions}). The impurity levels in this ultra-pure layer are tightly controlled to minimize spin decoherence:  $^{13}\text C + ^{14}\text C < 1$\thinspace ppb, boron $<$ 1\thinspace ppb, N $<$ 1\thinspace ppb.

At this stage, there are two possible paths forward (see Fig.\,4). In one approach, proposed by Marcus W. Doherty and co‑workers \cite{Oberginpress_bottom-up}, an atomically precise fabrication process is employed. This method combines hydrogen-desorption lithography (HDL) with scanning tunnelling microscopy (STM) to create and activate single NV centers with nanometer-scale positioning accuracy. Subsequently, a second ultra-pure diamond layer is grown on top of the NV-containing layer, so that the combined thickness of the two layers matches the intended height of the ND pillars. The resulting substrate, with NV centers embedded at predetermined locations, is then etched to form ND pillars (see Section \ref{Sec:fabrication}), each centered on an existing NV center.

In the second approach a delta-doped layer is grown on top of the ultra-pure single-crystal diamond layer---typically a 1–2\thinspace nm thick diamond layer incorporating nitrogen atoms (see, for example, \cite{Mobile_probe_nanodiamond_Awschalom,Jaffe2020NovelUltraLocalized}). We then grow a second, very thin (few nm) ultra-pure single-crystal diamond layer on top of the delta-doped layer.

\begin{wrapfigure}[17]{r}{0.45\textwidth}
  \vspace{0pt}  
  \raggedleft      
  \includegraphics[width=\linewidth]{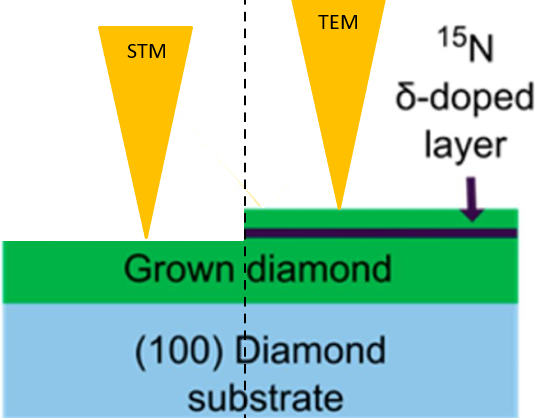}
  \caption{ Schematic STM and TEM procedures. Left:  STM in combination with HDL (see text) creates NV centers.   Right: The diamond is irradiated with a focused TEM electron beam to form localized vacancies. NV centers are formed after subsequent annealing (adapted from \cite {McLellan2016PatternedNV}).}
  \label{fig:TEM}
\end{wrapfigure}

In the delta-doped layer approach, our next step is to activate an x–y grid of single NV centers within the delta-doped layer. Here we  employ focused transmission electron microscopy (TEM) followed by annealing \cite{Jaffe2020NovelUltraLocalized,McLellan2016PatternedNV} (see Fig.\,\ref{fig:TEM}); the electrons  need to traverse only a few nanometers to reach the delta doped layer, resulting in negligible scattering and thus preserving spatial accuracy. The vertical positioning of the NV centers is inherently well controlled by the 1–2\,nm thickness of the delta-doped layer. Once activation is complete, we will finish growing the second ultra-pure single-crystal diamond layer, ensuring that the combined thickness of all three layers meets the target ND pillar height. As in the first approach, the resulting substrate---now containing NV centers embedded at predetermined locations---is subsequently etched to form ND pillars (see Sec.\,\ref{Sec:fabrication}), each precisely centered on an NV center.

In addition, we are invetigating a third option. This method utilizes a cold-ion gun with nanometer-scale precision \cite{Jacob2016DeterministicIonMicroscopy} to inplant nitrogen ion in predetermaind location and depth. Following implanetation and anneling we will grow the additional ultra pure diamond layer.

\section{Outlook}

\begin{figure}[htbp!]
  \centering
  \includegraphics[trim={30 230 30 90}, clip,width=0.9\linewidth]{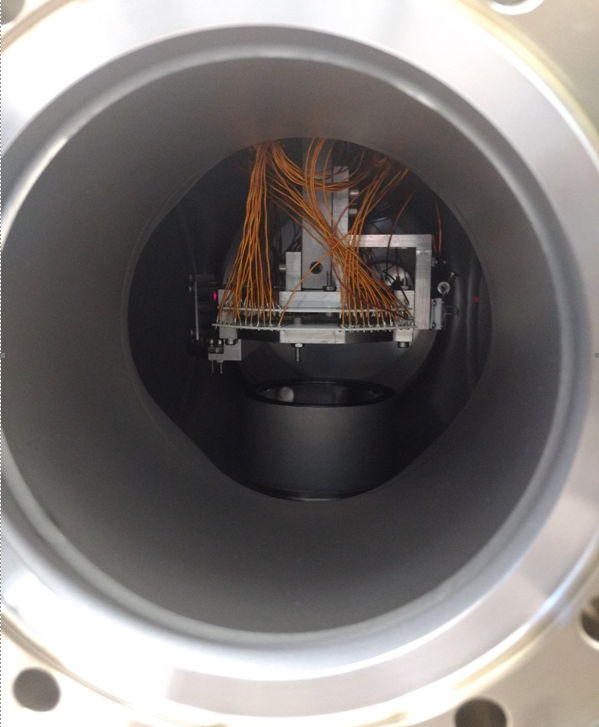}
  \caption {(Color online) The ultimate integrated chip for the ND SGI experiment. While the SGI experiment may be initially done with much more rudimentary setups, bringing together several independent elements each with its own function, we envision that eventually to increase accuracy all elements will be put on the same chip. For example, a planar Paul trap and current-carrying wires for a diamagnetic trap, should be perfectly aligned with the current-carrying wires delivering the ultra-high magnetic gradient for the SG forces. Depending on the results of our work on launching NDs \cite{Rafael_paper}, one of the integrated layers in this chip may also be the ND source described in this work. Alternatively, the source will be places elsewhere and the NDs will be delivered via an electric guide \cite{Rafael_paper}. An optimized structure is envisioned to include a double-layer chip with a gap of 2 mm where the trapping, cooling and interferometry take place. The bottom chip is made of transparent layers (i.e., Paul trap made of ITO) through which a high NA objective may manipulate the NV spin as well as detect the SGI outcome, while the top chip is made of special materials able to hold high current densities. One or both chips are also expected to hold a microwave antenna \cite{archive_of_Naor}. The double chip configuration has the additional advantage of covering most of the solid angle to which the ND is exposed, so that placing the chip on a cold finger would significantly suppress blackbody radiation. In this picture we show a preliminary double chip which we have fabricated and which has been tested in the labs of Ferdinand Schmidt-Kaler, where it successfully trapped cold beryllium ions  \cite{Wilzewski2020}.}
  \label{Fig:Double_Ion_chip}
\end{figure}

The purpose of the ND pillar source described in this note is to feed ND pillars into a Paul trap for the start of an interferometry cycle (see the beginning of Section \ref{Sec:Desing_consideration}). To accomplish this, the source must be integrated into a full experimental system. We note here that our group has a long and broad knowhow and experience integrating chips into experimental systems \cite{Keil2016Fifteen}. The source chip may eventually be positioned on its own at the far end of a loading guide\,\cite{Rafael_paper}, or, in rudimentary first versions of the ND SGI, may be directly adjacent to the trap position. The ultimate ND SGI chip is envisioned as a double chip with a 2\,mm gap in which the cooling and interferometry takes place (Fig.\,\ref{Fig:Double_Ion_chip}). The source may then become one of the layers in this double chip configuration.
To conclude, while we attempt to realize first versions of the ND SGI with the technology already available (e.g., commercially available NDs), we will continue to work in parallel on improving the available technology, and this includes a source for high-quality NDs. 

\section{Methods}
\label{Sec:Methods}
\subsection{Fabrication method of the ND$_1$ pillar}
\label{Sec:Fab_Methods}
In this subsection, we describe the fabrication procedure of the ND cuboidal pillars labeled  $\text{ND}_1$ (see Table \ref{Table:Dimentions}). The fabrication is carried out on a commercial single-crystal diamond substrate and serves to demonstrate the process capabilities, which will later be applied to the specialized multilayer diamond substrate (see section \ref{Sec:fabrication}).

We begin with a $2.6 \times 2.6\,$mm, $0.25\,$mm thick single-crystal chemical vapor deposition (CVD) diamond substrate (Element Six, SC plate CVD $2.6 \times 2.6 \times 0.25\,$mm, \textless100\textgreater, P2), oriented along the ⟨100⟩ crystallographic direction. The diamond is mounted on a $10 \times 10\,$mm, $0.75\,$mm thick p-type silicon substrate (boron-doped, resistivity $< 5\,\Omega \cdot \text{cm}$). Prior to processing, the silicon substrate is cleaned using a piranha solution, followed by thorough rinsing with deionized (DI) water. Subsequent megasonic cleaning at 3\thinspace MHz is carried out using an HMxSquare system (SUSS MicroTech). Final surface preparation includes a 10-minute oxygen plasma treatment in a Diener Plasma Asher to enhance adhesion and ensure complete cleaning. 

\begin{figure}[htbp]
  \centering
  \includegraphics[trim={70 480 50 45}, clip, width=0.95\linewidth]{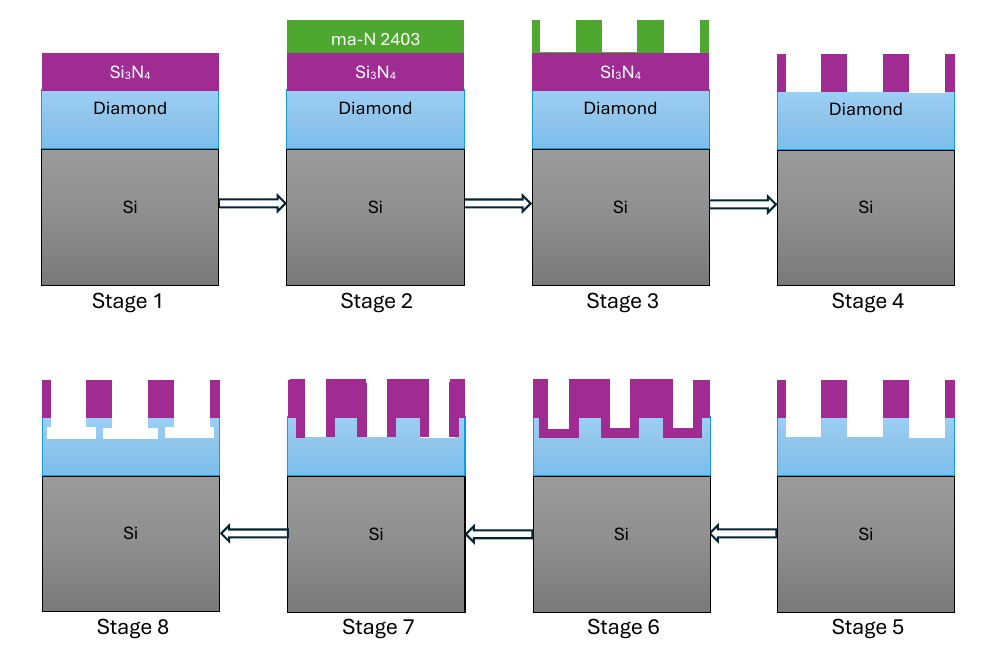} 
\caption{(Color online) Schematic illustration of the fabrication stages for ND$_1$ pillars (stages 1-5, details in Subsection~\ref{Sec:Fab_Methods}), including undercuts  (stages 6-8, details in Subsection~\ref{Sec:Undercut}). Stage 1 – Deposition of a 150\thinspace nm silicon nitride ($\text{Si}_3\text{N}_4$) layer using ICP-CVD. Stage 2 - spin coating of ma-N2403 resist. Stage 3 - lithography and development. Stage 4 - etching vertically the silicon nitride layer using DRIE. Stage 5 - etching vertically the exposed diamond. Stage 6 - deposition of thin protective silicon nitride layer. Stage 7 - etching vertically the silicon nitride layer. Stage 8 - isotropic etching, creating the undercuts.}
  \label{Fig:Undercut_Method}
\end{figure}

A 150\thinspace nm thick silicon nitride $(\text {Si}_3 \text N_4)$ layer is subsequently deposited (Fig. \ref{Fig:Undercut_Method}, stage 1) via inductively coupled plasma chemical vapor deposition (ICP-CVD) using an Oxford Instruments system, serving as a hard mask for etching. A high-resolution negative-tone electron beam resist (ma-N 2403, {M}icro {R}esist {T}echnology) is spin-coated at 3000\thinspace rpm and soft-baked at 90°C for 90 seconds (Fig. \ref{Fig:Undercut_Method}, stage 2). To suppress charging during electron beam lithography, a conductive coating (Electra 92, Allresist) was applied at 4000\thinspace rpm and baked at 90°C for 2 minutes.

Electron beam lithography is then performed using a Raith 5150 system (Fig. \ref{Fig:Undercut_Method}, stage 3) at an accelerating voltage of 100\thinspace kV, a beam current of 1\thinspace nA, and a dose of $1420\, \mu \text{C/cm}^2$. Arrays of rectangular features ($65 \times 45\,$nm, length × width) with a pitch of 1\thinspace $\mu$m are patterned over a $20 \times 20 \,\mu$m area. Additional rectangular features of varying sizes are also written for reference.  Following exposure, the sample is immersed in deionized (DI) water for 40 seconds to remove the conductive coating. The resist is then developed in AZ 726 MIF (2.38\% TMAH) for 8 seconds, followed by a 40-second DI rinse and then drying with nitrogen.

The underlying silicon nitride ($\text{Si}_3\text{N}_4$) hard mask is etched using deep reactive-ion etching (DRIE) in anisotropic (vertical) mode, using  an Oxford Cobra system for 37 seconds (Fig. \ref{Fig:Undercut_Method}, stage 4), with the following parameters: ICP power of 1000\thinspace W, RF forward power of 20\thinspace W, chamber pressure of 6\thinspace mTorr, and gas flow rates of 10\thinspace sccm $\text{SF}_6$ and 40\thinspace sccm $\text{CHF}_3$. Subsequently, the exposed diamond surface is etched in the same system and same mode (Fig. \ref{Fig:Undercut_Method}, stage 5) using an oxygen plasma for 40 seconds under the following conditions: ICP power of 850\thinspace W, RF forward power of 90\thinspace W, chamber pressure of 10\thinspace mTorr, and an $\text{O}_2$ flow rate of 30\thinspace sccm.

\subsection{Undercut fabrication method}
\label{Sec:Undercut}
In the previous Subsection (\ref{Sec:Fab_Methods}), we described the fabrication method developed for etching the ND$_1$ pillars. Here, we present the additional fabrication steps required to create undercuts that enable the separation of the NDs from the substrate. An undercut means that we create a narrow bottle neck where the pillar is connected to the substrate. This creates a brittle bond which could then be easily broken by vibrations from a transducer. The method involves depositing a sidewall protective layer, followed by quasi-isotropic etching.

After forming the ND pillars (Fig. \ref{Fig:Undercut_Method}, stage 5), we deposit a 100\thinspace nm silicon nitride sidewall protective layer (Fig. \ref{Fig:Undercut_Method}, stage 6) using ICP-CVD, as in stage 2. This thin film is then etched using deep reactive-ion etching (DRIE) in an Oxford Cobra system for 18 seconds under the following conditions: ICP power 1000\thinspace W, RF forward power 20\thinspace W, chamber pressure 6\thinspace mTorr, and gas flow rates of 10\thinspace sccm SF$_6$ and 40\thinspace sccm CHF$_3$. These parameters promote anisotropic (vertical) etching, removing material primarily from the top and bottom surfaces while leaving the sidewalls protected (Fig. \ref{Fig:Undercut_Method}, stage 7).


The subsequent quasi-isotropic etching step is performed in the ICP-CVD Oxford Instruments system under the following conditions: ICP power 1000\thinspace W, chamber pressure 20\thinspace mTorr, O$_2$ flow rate 30\thinspace sccm, table temperature $\sim 250^\circ\mathrm{C}$, and RF forward power 0\thinspace W. These parameters induce quasi-isotropic etching, forming the undercuts beneath the pillars while the silicon nitride layer protects the sidewalls (Fig. \ref{Fig:Undercut_Method}, stage 8). 

Silicon nitride removal is performed by immersing the sample in HF (Buffered Oxide Etch 6:1) for 2 minutes, followed by rinsing in deionized water and drying under a nitrogen flow.

\subsection{Fabrication process for ND$_2$ and ND$_3$ pillars}
\label{Sec:ND2_ND3}

In Section \ref{Sec:fabrication} and Subsection \ref{Sec:Fab_Methods}, we described the top-down fabrication processes used for producing the ND$_1$ pillars. Here, we outline the modifications required for fabricating the ND$_2$ and ND$_3$ pillars, which are primarily related to the significantly greater etch depths needed.

Based on process calibration, the etch selectivity of diamond to silicon nitride (used in the fabrication of ND$_1$ pillars, see Subsection \ref{Sec:Fab_Methods}) was determined to be $\sim 1.43$, while its selectivity to the ma-N 2400 series electron beam resist was $\sim 0.32$. Using these values, the required thicknesses of the silicon nitride hard mask for the ND$_2$ and ND$_3$ pillars etch depths were estimated to be $\sim 200\,$nm and $\sim 1.1\, $µm, respectively. Correspondingly, the necessary resist thicknesses were estimated at $\sim 650\,$nm for ND$_2$ pillars and $\sim 3.4\,$µm for ND$_3$ pillars.

While the ma-N 2400 series resist can provide film thicknesses up to approximately 1\thinspaceµm,---sufficient for ND$_2$ pillars, it is inadequate for ND$_3$ pillars. Therefore, to support both etch depths, a more robust resist system was adopted: the electron beam resist SX AR-N 8400.22 (Allresist), which offers high-resolution compatibility and can be spin-coated to a thickness of $\sim 400$\thinspace nm. Importantly, it demonstrates a selectivity of $\sim 2.7$ to silicon nitride \cite{Miakonkikh2021PlasmaResistanceHSQ}, making it suitable for patterning $\sim 1\,$µm of silicon nitride. 

This two-layer stack---SX AR-N resist over silicon nitride---enables effective pattern transfer for both ND$_2$ and ND$_3$ pillars. The SX AR-N 8400.22 process follows the same procedure described in Subsection \ref{Sec:Fab_Methods}, with appropriate adjustments to the exposure dose and baking temperature.

\subsection{Estimating the ND height}
\label{Sec:Height_Methods}

\begin{figure}[htbp!]
  \centering
  \includegraphics[trim={90 260 90 130}, clip,width=0.99\linewidth]{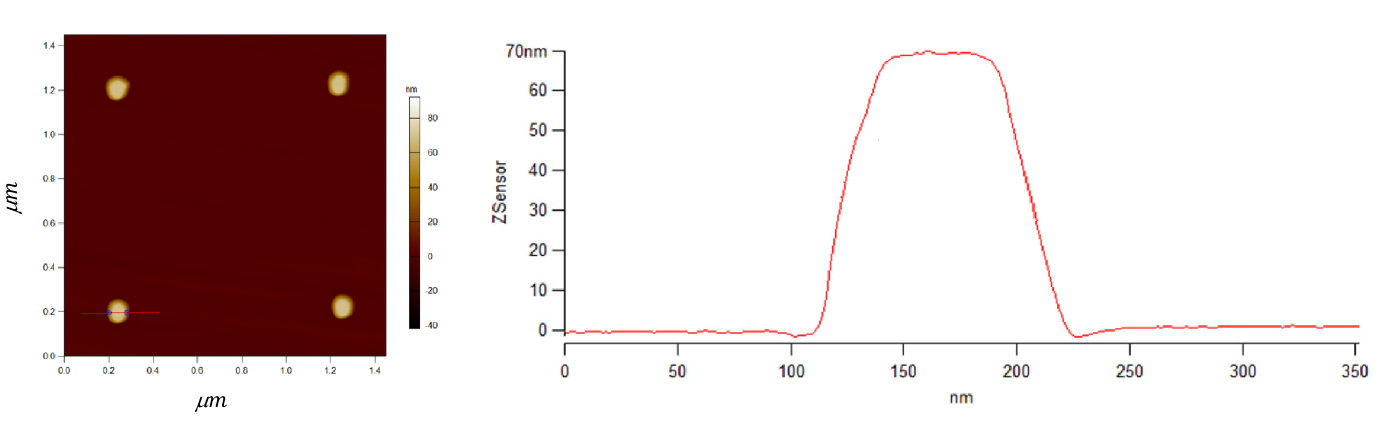}
  \caption {(Color online) AFM scanning results (1024 X 1024 points). Left: Height map obtained from an AFM scan over a $\sim 1.4 \times 1.4 \, \mu\text{m}$ region containing four ND pillars. Right: Line profile showing height as a function of lateral position along a cross-section through the center of the AFM image of the bottom-left ND pillar.}
  \label{Fig:AFM_1}
\end{figure}

The height of an ND pillar can be estimated from a tilted Scanning Electron Microscope (SEM) image; however, a more accurate measurement is obtained using an Atomic Force Microscope (AFM). In this approach, the same sample used for cross-sectional area estimation (see Subsection \ref{Sec:Measure_Methods}) is scanned with an MFP-3D-Bio AFM (Asylum Research, Oxford Instruments). Imaging is performed in AC-mode using an AC160TS probe (Olympus). Instrument control and data analysis are conducted using Igor Pro software (WaveMetrics). Figure \ref{Fig:AFM_1} shows a height map of a $\sim 1.4 \times 1.4 \, \mu\text{m}$ region containing four ND pillars and the corresponding height profile along a lateral cross-section through the center of one of the pillars. The profile reveals that the base region is flat within 1–2\thinspace nm, and the top surface of the pillar is also uniformly flat. Analysis of a similar height map covering 25 ND pillars confirms that the base area remains flat within 1–2 nm and reveals an average pillar height of 68.2\thinspace nm with a standard deviation of 0.5\thinspace nm, or relative standard deviation of 0.7\%

\subsection{Estimating the ND pillar volume and mass}
\label{Sec:Measure_Methods}

One of the design requirements (see Section \ref{Sec:Desing_consideration}) is the uniformity of the ND pillar's mass, or equivalently, their volume. The pillars have a cuboidal shape, and their volume can be estimated by multiplying their height (discussed in Subsection \ref{Sec:Height_Methods}) by their cross-sectional area (i.e., length x width).  The measurement of linear nanoscale dimensions using SEM has been extensively reviewed  (\cite{Crouzier2019Methodology_SEM_measure,Paras_Low-Dim_nanomaterials,Bertness2017Dimensional_with_SEM,Feltin_SEM_Meas_nanoscale} and references therein). Based on these publications and the authors’ experience, using SEM to measure features of approximately 50\thinspace nm in length results in an expected uncertainty of at least $\pm2\,$nm, ($\pm 4\%$), which translates to an uncertainty of approximately $\pm 8\%$ in the estimated cross-sectional area. 

This uncertainty can be mitigated by using direct area measurement tools such as \textit{ImageJ} \cite{schneider2012nih_ImageJ}. Such programs segment the object from the background, trace its perimeter, and estimate its area by counting the number of enclosed pixels. Naturally, as the ratio between the object’s area and the pixel size increases, the variance introduced by this estimation method decreases. It is worth noting that, in our case, the primary concern is the uniformity of ND volumes fabricated on a single substrate, rather than variations between different substrates. Consequently, our measurement approach does not require strict traceability to absolute physical standards.
 
The analyzed ND pillars have a rectangular cross-section, defined by the etching of a masked diamond substrate, as described in Subsection \ref{Sec:Fab_Methods}. Their nominal length and width values are listed in Table \ref{Table:Dimentions}. SEM characterization was performed using a JEOL JSM-IT800 SHLs Schottky Field Emission SEM (JEOL Ltd.) at magnifications of 33k×–60k×. The system operated in Super Hybrid Lens (SHL) mode with a low accelerating voltage of 1.00 kV and a working distance of 2.9 mm. Images were acquired using the Upper Hybrid Detector (UHD), which simultaneously collects secondary electrons (SE) and backscattered electrons (BSE) to provide high-contrast, detailed images. This configuration allows SEM images with a high signal-to-noise ratio (S/N) to be obtained even at very low accelerating voltages, enabling the resolution of fine nanoscale surface features.

The SHL mode, which combines electrostatic and magnetic lens fields, enhances optical performance and spatial resolution. Low-voltage operation reduces electron beam penetration into the diamond, increasing surface sensitivity and minimizing charging artifacts—an important advantage for imaging insulating materials and delicate nanostructures.

\begin{figure}[htbp]
  \centering
  \includegraphics[trim={20 400 20 20}, clip, width=0.95\linewidth]{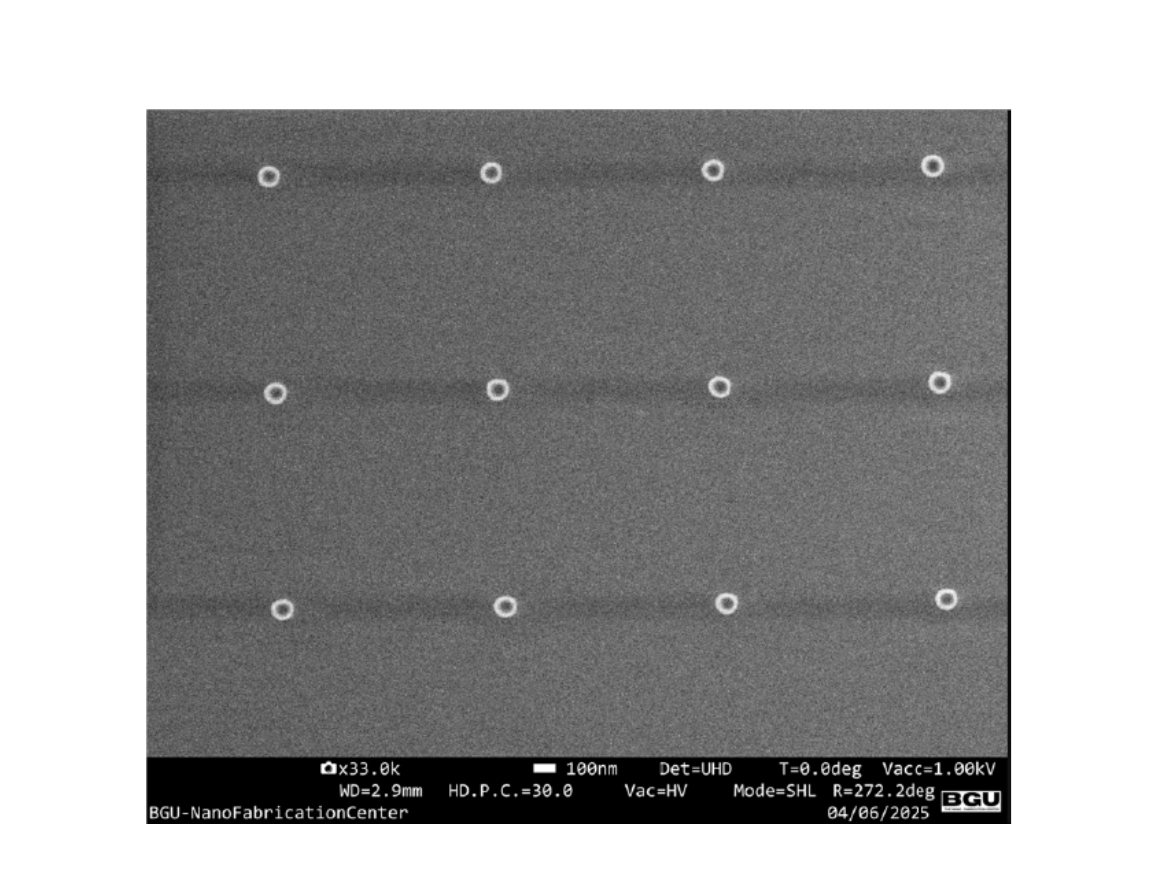} 
  \caption {(Color online) A typical SEM image used to estimate NDs' volume  homofginity. Shown is one of $N$ SEM top view images, displaying 12 ND pillars (bright rings) after gold deposition (see text).   The imaging resolution is 505\thinspace pm/pixel.}
  \label{Fig:SEM_Method}
\end{figure}

Due to the diamond’s insulating nature, electrical charging can blur feature boundaries in SEM images. To mitigate this effect, a 5 nm gold layer was deposited on the sample surface using an AJA Orion-8-UHV sputter system (AJA International), improving contrast between the features and the background. Figure \ref{Fig:SEM_Method} shows an example SEM image of 12 ND pillars after gold deposition, with projected dimensions of $40 \times 65$ nm and an imaging resolution of 505 pm/pixel.

The area estimation procedure is based on the analysis of top-view SEM images of the ND pillars, processed using \textit{ImageJ} \cite{schneider2012nih_ImageJ}, an open-source and public-domain image analysis software. In a typical dataset, $N$ SEM images are analyzed (e.g., Fig. \ref{Fig:SEM_Method}), each containing $M$ distinguishable ND pillars. Prior to analysis, the spatial calibration of each SEM image is updated in \textit{ImageJ} based on scale information embedded in the SEM metadata. .  Subsequently, the operator applies a manual grayscale threshold to segment the ND pillars from the background. The software then computes and returns the estimated areas of each ot  the ND pillars within each image.  The estimated area of the $m^{\text{th}}$ pillar in the $n^{\text{th}}$ image is denoted $S_{nm}$

In our imaging protocol, the number of ND pillars ($M$) captured in each frame is deliberately limited to facilitate the use of high magnification and correspondingly small pixel sizes. Thus, many images ($N$) must be analyzed to obtain reliable statistics. However, the grayscale threshold selected by the operator for each image significantly affects the computed values of $S_{nm}$,  introducing additional variance. To mitigate this effect, a normalization procedure is employed. First, a local average projected area is computed for each image:

\begin{equation}
\text{SA}_n = \frac{1}{M} \sum_{m=1}^{M} S_{nm}
\label{Eq:local_average}
\end{equation}

Next, a global average projected area is computed over all images:

\begin{equation}
\text{SAA} = \frac{1}{N} \sum_{n=1}^{N} \text{SA}_n
\end{equation}

Finally, each individual measurement ${S_{nm}}$ is normalized using these averages:

\begin{equation}
\overline{S}_{nm} = \frac{\text{SAA}}{\text{SA}_n} \, S_{nm}
\end{equation}

This normalization reduces the influence of threshold variability across different images and improves the consistency of the area estimates. Additional "in-image" normalization may also be required if the background grayscale varies significantly within a single SEM image, introducing further variation in the estimated area of ND pillars located in different regions. To mitigate this effect, the operator manually adjusts the grayscale threshold in different parts of the image, thereby achieving a form of "double normalization".

\subsection{Fabrication of a high-resolution master target on a silicon substrate}
\label{Sec:Master_target}

Nanometric master targets are precision-engineered reference structures—often periodic gratings or patterned thin‑films—with dimensions calibrated against national standards. They serve as critical benchmarks for traceable calibration and uncertainty reduction in nanoscale metrology, including SEM image analysis (see, for exapmle \cite{Ukraintsev2012}). A recent study \cite{Feltin_SEM_Meas_nanoscale} used a Certified Reference Material (CRM) consisting of silica nanoparticles, with a certified equivalent circular diameter of $83.7 \pm 2.2$\thinspace nm, randomly deposited on a substrate. However, we were unable to identify a master target suitable for validating our cross‑sectional area estimation method (see Subsection \ref{Sec:Measure_Methods}) for our regularly spaced nanopillars of type ND$_1$. Consequently, we fabricated a dedicated master target for this purpose, as detailed below. While this master target is not traceable to physical standards and therefore cannot be used to establish absolute accuracy, it may be adequate for evaluating the variance introduced by our method when assessing in‑substrate uniformity.

This high-resolution master target was fabricated on a silicon substrate using electron beam lithography (EBL) and thin-film deposition techniques. The substrate was a p-type boron doped silicon, with a resistivity  $< 5\,\Omega \cdot \text{cm}$ and a thickness of $\sim 750\,\mu \text m$. The process began with substrate cleaning using a piranha solution, followed by thorough rinsing with deionized (DI) water using a HMxSquare SUSS MicroTech system to remove organic residues. To further improve surface cleanliness and resist adhesion, the substrate was treated with oxygen plasma for 10 minutes using a Diener Plasma Asher. A high-resolution negative-tone resist, ma-N 2403 ({M}icro {R}esist {T}echnology) was diluted to $\sim 160\, \text{nm}$ and  spin-coated at 3000\thinspace rpm. Prebaking was performed at 90°C for 90\thinspace sec. Patterning was carried out using a Raith 5150 EBL system at 100\thinspace kV and 1\thinspace nA, $200\,\mu\text{A}$ with a dose of $1400\, \mu \text{C/cm}^2$. Arrays of circular features, each 60\thinspace nm in diameter with a 500\thinspace nm pitch, were written over a $500 \times 500\,$µm area to define the resolution target layout. After exposure, the resist was developed in AZ 726 MIF developer (2.38\% TMAH) for 40 seconds, followed by a 40-second DI water rinse. A ~5\thinspace nm gold layer was then deposited using an AJA Orion-8 sputter coater to create visible and robust gold-capped pillars. The gold coating was applied primarily to improve image quality during scanning electron microscope (SEM) observation by enhancing surface conductivity and reducing charging effects, which in turn increases image contrast and resolution. The resulting structures consist of resist pillars with $\sim 160\,$nm height, serving as a nanoscale resolution test pattern.

\subsection{Measuring method validation}
\label{Sec:validation}

We use SEM images in combination with \textit{ImageJ} analysis software  (see Subsection \ref{Sec:Measure_Methods})  to estimate the area of the fabricated NDs (see Subsection \ref{Sec:Fab_Methods} and \ref{Sec:ND2_ND3})  as well as the structures on the fabricated master target (see Subsection \ref{Sec:Master_target}).  
 
\begin{figure}[htbp!]
  \centering
  \includegraphics[trim={0 0 0 0}, clip,width=0.6\linewidth]{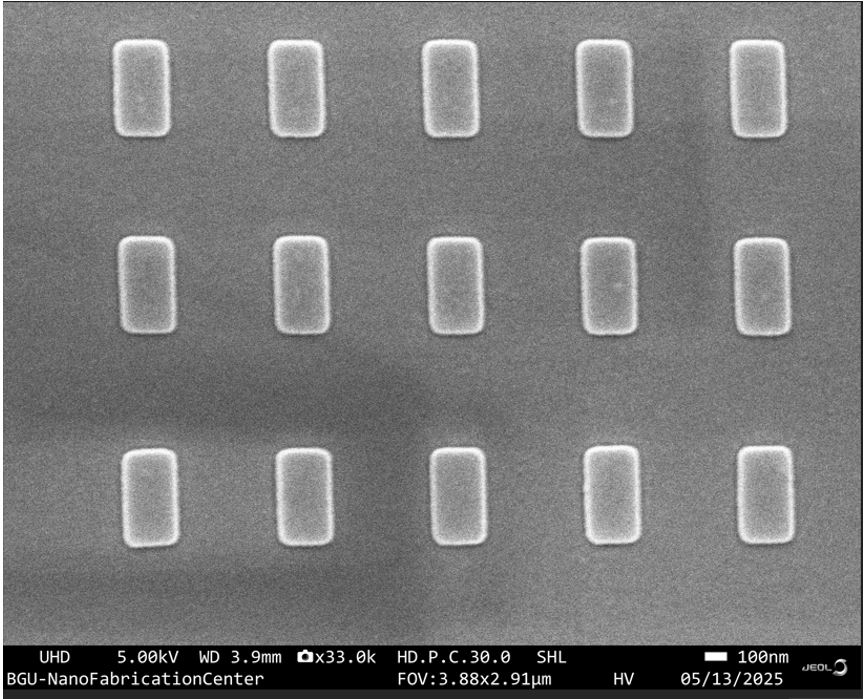}
  \caption {(Color online) Top-view SEM image showing a group of 15 nanodiamond pillars, each with nominal dimensions of $250 \times 400\,$nm, as part of a larger pillar array.}
  \label{Fig:large_structures}
\end{figure}

Our primary focus is on the uniformity of the ND pillars rather than their absolute size (see the end of Section \ref{Sec:Desing_consideration}). For relatively large structures (see Fig. \ref{Fig:large_structures}), we obtain (using the method defined in Subsection \ref{Sec:Measure_Methods}) a relative standard deviation (RSD) of 1\% across a group of 60 nanopillars. The RSD (defined as the standard deviation divided by the mean and expressed as a percentage) has two main sources: (i) inherent variations in the actual pillars' areas, and (ii) additional variance introduced by SEM imaging and the area estimation process using \textit{ImageJ}. Clearly, in this case, as the total variation is 1\%, the contribution of the measurement method to the total variance must be below 1\%.

\begin{figure}[htbp!]
  \centering
  \includegraphics[trim={0 0 0 0}, clip,width=0.75\linewidth]{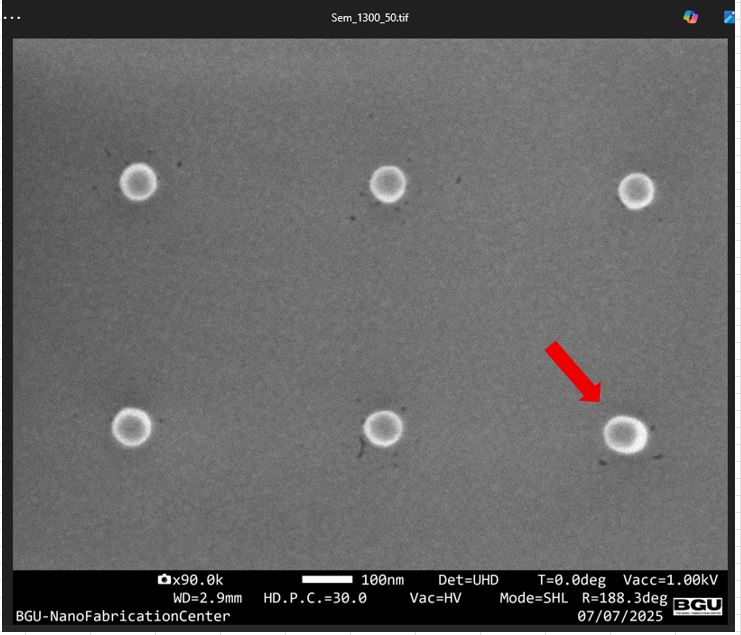}
  \caption {(Color online) Representative SEM image of the master target.  Shown are six structures. One structure, marked with a red arrow, is noticeably larger than the others.}
  \label{Fig:master_target}
\end{figure}

When we measure smaller structures---where the ratio between the object’s area and the pixel area is lower---the measuring method will introduce greater variation. Our smallest structures are approximately $40 \times 65\,$nm in size. To investigate the variation introduced by our measurement procedure, we prepared a master target with round structures 60\thinspace nm in diamter  (see Subsection \ref{Sec:Master_target}) and applied the same measurement protocol (see Subsection \ref{Sec:Measure_Methods}) to it. This yielded a dataset of the measured areas for these structures.

A representative SEM image of the master target, which contains six structures, is shown in Fig.\,\ref{Fig:master_target}. One of these structures is clearly different from the others, and similar outlier structures were observed in additional images. Consequently, the measurement protocol (Subsection \ref{Sec:Measure_Methods}) was applied twice: first to the full set of 33 SEM images (each containing six structures) and then to the same 33 images after removing the outlier structures. The resulting relative standard deviations of the measured areas were 6.6\% in the first case and 3.1\% in the second.   

In addition, AFM images (see Subsection \ref{Sec:Height_Methods}) were used to estimate the areas of 24 structures on the master target. This was done by counting the number of points above a certain threshold for each structure. The results were compared to those obtained using \textit{ImageJ} analysis of the SEM images of the same structures. We found that the two series had similar relative standard deviations (6.5\% and 5.6\%, respectively). The Pearson correlation coefficient (\cite{Kenton2023Pearson}, see also \cite{WikipediaPearson2025}) was found to be 0.08, indicating no correlation between the two methods. 

A dedicated study is currently underway to investigate the origins of this discrepancy, with the goal of refining both measurement techniques to achieve improved accuracy and consistency.

\begin{acknowledgments}
We thank the BGU Atom-Chip Group support team, especially  Zina Binstock and Dmitrii Kapusta for their support in building and maintaining the experiment. Funding: This work was funded by the Gordon and Betty Moore Foundation (doi.org/10.37807/GBMF11936) and by the Simons Foundation (MP-TMPS-00005477). Additional funding for advanced fabrication was provided by the Israel Science Foundation, grant {N}o. 3470/21.
\end{acknowledgments}

\clearpage

\end{document}